\documentstyle[12pt,aps,prd,preprint]{revtex}
\begin{document}
\title{Black-Hole Polarization and Cosmic Censorship}
\author{Shahar Hod}
\address{The Racah Institute for Physics, The
Hebrew University, Jerusalem 91904, Israel}
\date{\today}
\maketitle

\begin{abstract}
  The destruction of the black-hole event horizon is ruled out by both
  cosmic censorship and the generalized second law of thermodynamics.
  We test the consistency of this prediction in a (more) `dangerous'
  version of the gedanken experiment suggested by Bekenstein and
  Rosenzweig. A $U(1)$-charged particle is lowered {\it slowly} into a
  near extremal black hole which is not endowed with a $U(1)$ gauge
  field. The energy delivered to the black hole can be {\it
    red-shifted} by letting the assimilation point approach the
  black-hole horizon. At first sight, therefore, the particle is not
  hindered from entering the black hole and removing its horizon.
  However, we show that this dangerous situation is excluded by a
  combination of {\it two} factors not considered in former gedanken
  experiments: the effect of the spacetime curvature on the
  electrostatic {\it self-interaction} of the charged system (the
  black-hole polarization), and the {\it finite} size of the charged
  body.

\end{abstract}
\bigskip

The {\it cosmic censorship} hypothesis, introduced by Penrose
\cite{Pen} thirty years ago, is one of the corner stones of general
relativity. Moreover, it is being envisaged as a basic principle of
nature. The validity of this conjecture is, however, still an
open question in classical general relativity. 

The destruction of a black hole (i.e., its event horizon) is ruled out by this
principle because that would expose naked singularities to distant
observers. Moreover, the horizon area $A$ of a black hole is associated with
entropy $S_{bh}=A/4$ (we use gravitational units in which $G=c=1$). 
Therefore, without any obvious physical mechanism to compensate for the
loss of (black-hole) entropy, 
the destruction of the black-hole event horizon is expected
to violate the generalized second law of thermodynamics
\cite{Beken1}. For these two reasons, any process which seems, at
first sight, to have a chance of removing the black-hole 
horizon is expected to be unphysical. For the advocates of the cosmic
censorship principle the task remains to find out how such candidate
processes eventually fail to remove the horizon.

In this paper we inquire into the physical mechanism which protects
the black-hole horizon from being eliminated by the capture of a
charged particle which ``supersaturate'' the extremality condition for
black holes. As is well known, the Kerr-Newman metric 
with $M^2 < Q^2 +a^2$ (where $M,Q$ and $a$ are the mass, charge, and angular momentum
per unit mass of the configuration) does {\it not} contain an event
horizon, and therefore describes a naked singularity.

We begin with the type of gedanken experiment suggested by Bekenstein
and Rosenzweig \cite{BekRos} and independently by Hiscock \cite{His} 
(our version of the gedanken experiment
is, however, more `dangerous' than the one considered in
\cite{BekRos}): suppose there exist two different types
of local charge, namely type-$q \in U(1)$ and type $k \in U'(1)$, e.g.,
electric and magnetic charge. A Reissner-Nordstr\"om spacetime with {\it two} different 
charges $Q \in U(1)$ and $K \in U'(1)$ displays an event horizon 
only if $Q^2 +K^2 \leq M^2$. 

The black hole is assumed to be a near
extremal Reissner-Nordstr\"om black hole, possessing a $U'(1)$
charge $K$, but {\it no} $U(1)$ charge. Thus, the black hole is not
endowed with a $U(1)$ gauge field, and an infalling charge $q$
[where $q \in U(1)$] seems to encounter no repulsive electrostatic potential
barrier. Therefore, at first sight, the particle is not forbidden from
crossing (and removing) the black-hole horizon. An assimilation of 
a charged body with proper energy (energy-at-infinity) $E$, and 
charge $q$, results with a 
change $dM=E$ in the black-hole mass and a change $dQ=q$ in its
charge; Thus, a necessary 
condition for removal of the horizon after the assimilation of
the body is $E \leq {(q^2 +K^2-M^2)}/{2M} \leq q^2/2M$.

Bekenstein and Rosenzweig \cite{BekRos} considered the infall of the charged
particle from spatial {\it infinity}. For this case $E \geq \mu$,
where $\mu$ is the particle's rest mass, and
a ({\it necessary}) condition for removal of the horizon is therefore 

\begin{equation}\label{Eq1}
q^2/\mu \geq 2M\  .
\end{equation}
This condition simply says that the classical radius of the charged
body (the analogous of the well-known classical radius of the
electron) is larger than the black-hole size. Thus, if the body is capable of
fitting in the black hole, it {\it cannot} satisfy condition
(\ref{Eq1}), and thus cannot remove the horizon.

A question immediately arises: what physical mechanism insures the
stability of the horizon if the charged particle is {\it lowered} slowly
towards the black hole ? In this case the energy delivered
to the black hole can be arbitrarily {\it red-shifted} by letting the
assimilation point approach the black-hole horizon. Therefore, this
type of gedanken experiment is more `dangerous' than the one
considered in \cite{BekRos}. At first sight,
therefore, the particle is not hindered from entering the black hole and removing
its horizon. However, in this paper we shall show that the dangerous
particle fails to remove the horizon; this conclusion 
is a direct consequence of {\it two} factors
not considered in \cite{BekRos}: the black-hole polarization, and 
the {\it finite} size of the charged body.

We consider a charged body of rest mass $\mu$, charge $q$, and proper
radius $b$, which is lowered towards a (near extremal) black hole. 
The total energy $E$ of a body moving in a black-hole spacetime 
is made up of the energy $E_0$ of the body's mass
(red-shifted by the gravitational field), and the
electrostatic self-energy $E_{self}$ of the charged body 
(there is no repulsive electrostatic force between the
charges $K$ of the black hole and $q$ of the body, since these charges
belong to {\it different} $U(1)$ gauge fields).

The first contribution, $E_0$, is given by Carter's \cite{Carter}
integrals of the Lorentz equations of motion of a charged particle
moving in a black-hole background \cite{note1}:

\begin{equation}\label{Eq2}
E_0={{\mu (r_{+}-r_{-})^{1/2}} \over
  {r_{+}}} \delta^{1/2}\left\{1+
O\left[{\delta /(r_{+}-r_{-})}\right] \right\}\  ,
\end{equation}
where $\delta = r-r_+$, and $r_{ \pm}=M \pm (M^2-K^2)^{1/2}$ are the
locations of the black-hole (event and inner) horizons.

The gradual approach to the black hole must stop when the
proper distance from the body's center of mass to the black-hole
horizon equals $b$, the body's radius. Thus, one should
evaluate $E_0$ at $r=r_{+}+ \delta (b)$, where $\delta(b)$ is 
determined by 

\begin{equation}\label{Eq3}
\int_{r_{+}}^{r_{+}+ \delta (b)} (g_{rr})^{1/2} dr = b\  ,
\end{equation}
with $g_{rr}=r^2/(r-r_+)(r-r_-)$. 
Integrating Eq. (\ref{Eq3}) one finds (for $b \ll r_{+}$)

\begin{equation}\label{Eq4}
\delta (b)=(r_{+}-r_{-}) {b^2/{4{r_{+}}^2}}\  .
\end{equation}
Thus, we obtain $E_0=\mu b (r_+-r_-)/2{r_+}^2$.

The second contribution, $E_{self}$, reflects the effect of the 
spacetime {\it curvature} on the particle's 
electrostatic {\it self-interaction}. The physical origin of this force 
is the distortion of the charge's long-range Coulomb field by
the spacetime curvature. This can also be interpreted as being due to the
image charge induced inside the (polarized) black hole
\cite{Linet,BekMay}. The self-interaction of a charged particle in the
black-hole background results with a {\it repulsive} (i.e., directed away from the
black hole) self-force. A variety of
techniques have been used to demonstrate this effect in black-hole 
spacetimes \cite{DeDe,Ber,Mac,Vil,SmWi,ZelFro,Lohi,LeLi1,LeLi2}. In particular, 
the contribution of this effect to the
particle's (self) energy in the Reissner-Nordstr\"om 
background is $E_{self}=Mq^2/2r^2$ \cite{ZelFro,Lohi}, which implies
$E_{self}=Mq^2/2{r_+}^2$ to leading order in $(b/r_+)^2$.

An assimilation of the charged body results with a change $dM=E$ in the
black-hole mass and a change $dQ=q$ in its $Q$-type charge. The condition
for the black hole to preserve its integrity after the assimilation of
the body is 

\begin{equation}\label{Eq5}
q^2 +K^2 \leq (M+E)^2\  .
\end{equation}
Substituting $E=E_0+E_{self}$ we find that a necessary condition for
removal of the black-hole horizon is

\begin{equation}\label{Eq6}
\mu < {q^2 \over b} -{{(r_+-r_-) ({r_+}^2-q^2)}\over {4M b}} <{q^2 \over b} \  .
\end{equation}

The total mass of the charged body is given by $\mu =\mu_0+ fq^2/b$,
where $\mu_0$ is the mechanical (nonelectromagnetic) mass, and $f$ is
a numerical factor of order unity which depends on how the charge is
distributed inside the body. The Coulomb energy 
attains its {\it minimum}, $q^2/2b$, when the charge is uniformly spread on a thin
shell of radius $b$, which implies $f \geq 1/2$ (an homogeneous
charged sphere, for instance, has $f=3/5$). Therefore, any charged body which respects the
weak (positive) energy condition must be larger than $r_c \equiv
q^2/2\mu$. 

In deriving the lower bound on particle's size $r_c$ 
one neglects the mechanical mass of
the body. In fact, large stresses may be placed inside the charged
body and the charge distribution must have forces of nonelectromagnetic
character holding it stable.
A purely classical electromagnetic model therefore has little
relevance to the {\it real} world. Nonelectromagnetic forces imply a 
large contribution $\mu_0$ to the mass of the body from such forces. The
large nonelectromagnetic contribution will
prevent us from coming close to the minimum size limit $r_c$: Atomic
nuclei, for instance, are bounded by {\it strong} forces,
which are often much stronger than the force exerted by the surface
electric field. In fact, even atomic nuclei, which are 
the {\it densest} charged objects (with negligible self-gravity) in
nature, satisfy the relation $b/{r_c} \sim
10^2-10^3$ and are therefore {\it far larger} than $r_c$ ! Black holes
with their extreme {\it gravitational} binding character are in fact the only
objects in nature whose size can come close to the limit $r_c$: An 
extremal Reissner-Nordstr\"om black hole, in particular, 
satisfies the relation $b/{r_c}=2$ (other black holes satisfy
$b/{r_c}>2$). Therefore, even the gravitational interaction in its
extremal form as displayed in black holes {\it cannot} allow a charged
object to be as small as $r_c$. 

Thus, one may safely conjecture that a charged body which respects the
weak (positive) energy condition
must satisfy the restriction $b/{r_c} \geq 2$ (where the equality
is only saturated by the extremal Reissner-Nordstr\"om black hole),
and we therefore conclude that the black-hole horizon cannot be
removed by an assimilation of such a charged body -- cosmic
censorship is upheld !

For an elementary charge which is subjected to Heisenberg's quantum
uncertainty principle with $b \sim \hbar/ \mu$, 
$r_c$ is not the measure of particle size. In fact, for $U(1)$ charges found free in nature
(weak coupling constant $q^2 \ll \hbar$, e.g., an electron), the classical radius $r_c$
is far smaller than the Compton length. This is incompatible with the necessary
condition Eq. (\ref{Eq6}), and we therefore recover our
previous conclusion that the black-hole horizon cannot be removed. 

The gedanken experiment can be generalized to astrophysically
realistic black holes, i.e., to {\it rotating} Kerr black holes. 
We consider a charged body which is lowered 
towards the black hole along its symmetry axis. 
For this case one finds
$\delta(b)=(r_{+}-r_{-}){b^2/{4({r_+}^2+a^2)}}$, 
where $r_{ \pm}=M \pm (M^2-a^2)^{1/2}$. This 
yields (see e.g., \cite{Beken2}) $E_0=\mu b (r_+-r_-)/2(r_+^2+a^2)$. 
The self-interaction $E_{self}$ of a charged particle on the symmetry
axis of a Kerr black hole has been determined in \cite{Lohi,LeLi1}, 
yielding $E_{self}=Mq^2/2(r^2+a^2)$. The assimilation of the charged
particle by the black hole produces the following changes in the
black-hole parameters:

\begin{equation}\label{Eq7}
M \to M+E\ \ \ ;\ \ \ a \to a[1-2E/M+O(E^2/M^2)]\ \ \ ;\ \ \ Q=0 \to
q\  .
\end{equation}
Hence, the condition for the black hole to preserve its integrity after the assimilation of
the body 
is (to leading order in $E/M$) $q^2 + a^2(1-E/M) \leq M^2(1+2E/M)$. 
Substituting $E=E_0+E_{self}$ we find that a necessary condition for
removal of the black-hole horizon is

\begin{equation}\label{Eq8}
b < {q^2 \over \mu} {M^2 \over {M^2+a^2}}-
{{M(r_+-r_-)({r_+}^2+a^2-q^2)} \over {4(M^2+a^2) \mu}} <{q^2 \over \mu}\  .
\end{equation}
This is the same as condition (\ref{Eq6}), and we therefore recover
our previous conclusion that classical charged bodies which satisfy the
weak energy condition, and elementary particles which are subjected to
Heisenberg's quantum uncertainty principle 
cannot remove the black-hole horizon -- the cosmic
censorship conjecture is respected !

The synthesis of these two gedanken experiments [namely, the capture of
a particle with $q \in U(1)$ charge by a Kerr-Newman black hole of
charge $K \in U'(1)$] is trivial. We obtain the same condition
(\ref{Eq8}) for the destruction of the black-hole horizon, where 
now $r_{ \pm}=M \pm (M^2-K^2-a^2)^{1/2}$. This implies, again, that
the horizon cannot be removed in our `dangerous' gedanken experiment.

Evidently, Eqs. (\ref{Eq6}) and (\ref{Eq8}) implies that not only
processes which transcend the extremality condition are forbidden, but
also processes which saturate this condition are forbidden as
well. In other words, a (near extremal) black hole {\it cannot} be
transformed into an extremal one by assimilating a charged
particle. This conclusion generalizes the results of \cite{Wang} to
the gedanken experiments considered in this paper. 

In summary, we have tested the consistency of the cosmic censorship
conjecture and the generalized second law of thermodynamics in the
simple context of a charged particle lowered into a (near extremal) 
black hole. In particular, we have considered a (more) `dangerous'
version of the gedanken experiment
suggested by Bekenstein and Rosenzweig \cite{BekRos}: The particle is
lowered {\it slowly} into the black hole and the 
energy delivered to the 
black hole is, therefore, {\it red-shifted}. Hence, at 
first sight, the particle is not hindered from entering the black hole and removing
its horizon. We have shown, however, that the black-hole horizon is
protected by a combination of {\it two} factors not considered in former 
gedanken experiments: the influence of the spacetime curvature 
on the electrostatic {\it self-interaction} of the charged body, 
and the {\it finite} size of the charged system. 

\bigskip
\noindent
{\bf ACKNOWLEDGMENTS}
\bigskip

I thank Jacob D. Bekenstein and Avraham E. Mayo for helpful discussions. 
This research was supported by a grant from the Israel Science Foundation.

\appendix
\def\q{q}
\section{Lowering an elementary charge with $\q^2>\hbar$}

Throughout this paper we have assumed that the charged body obeys the
weak (positive) energy condition. However, an elementary charge with
$b \sim \hbar/\mu$ in a
strongly coupled QED-type theory (with $q^2>\hbar$), has $b < q^2/\mu$,
and we cannot rule out condition (\ref{Eq8}). Nevertheless, Bekenstein
and Rosenzweig \cite{BekRos} have shown that for a quantum charge the
horizon is saved because in order to avoid the Landau ghost, the
effective coupling constant cannot be large enough to accomplish the
removal of the horizon (see \cite{BekRos} for a detailed analysis). 
Actually, the analysis of \cite{BekRos} concerning 
{\it quantum} particles is also applicable to our case, in which the
particle is lowered {\it slowly} into the black hole. In particular, 
Eq. (19) of \cite{BekRos} still holds in our case:

\begin{equation}\label{EqA1}
[q(\zeta M)]^2 <3\pi \hbar /2\ln(\zeta \xi M \mu / \hbar)\  ,
\end{equation}
with $\zeta >1$ and $\xi >1$ (see again \cite{BekRos} for details). 
The necessary condition for removal of
the black-hole horizon $b<q^2/\mu$ 
implies $\mu b \ln(\zeta \xi M \mu / \hbar) <3\pi \hbar /2$. 
However, since $\zeta \xi$ must be few times unity, this
condition can be satisfied only if $b=O(M)$. The motion of a particle
of this size in the black-hole background [which has a characteristic
length scale of $O(M)$] cannot be treated classically; its evolution in
the black-hole background is quantum mechanical in its
nature. Therefore, we cannot conclude that the horizon can be removed
by these particles \cite{BekRos}.


\begin{thebibliography}{99}

\bibitem{Pen} R. Penrose, Riv. Nuovo Cimento {\bf 1}, 252 (1969).

\bibitem{Beken1} J. D. Bekenstein, Phys. Rev. D {\bf 9}, 3292 (1974).

\bibitem{BekRos} J. D. Bekenstein and 
C. Rosenzweig, Phys. Rev. D {\bf 50}, 7239 (1994).

\bibitem{His} W. A. Hiscock, Ann. Phys. (N. Y.) {\bf 131}, 245 (1981).

\bibitem{Carter} B. Carter, Phys. Rev. {\bf 174}, 1559 (1968).

\bibitem{note1} Our task is to challenge the validity of the 
cosmic censorship conjecture in 
the most `dangerous' situation, i.e., when the charge-to-energy ratio of
the particle is as large as possible. Therefore, we consider a body which is 
captured from a radial turning point of its motion. 
This {\it minimize} the energy delivered to the
black hole (for a given charge of the body).

\bibitem{Linet} B. Linet, J. Phys. A: Math. Gen. {\bf 9}, 1081 (1976).

\bibitem{BekMay} J. D. Bekenstein and A. E. Mayo, e-print gr-qc/9903002.

\bibitem{DeDe} C. M. DeWitt and B. S. DeWitt, Physics (N. Y.) {\bf 1}, 3 (1964).

\bibitem{Ber} F. A. Berends and R. Gastmans, Ann. Phys. (N. Y.) {\bf 98}, 225 (1976).

\bibitem{Mac} C. H. MacGruder III, Nature (London) {\bf 272}, 806 (1978).

\bibitem{Vil} A. Vilenkin, Phys. Rev. D {\bf 20}, 373 (1979).

\bibitem{SmWi} A. G. Smith and C. M. Will, Phys. Rev. D {\bf 22}, 1276
  (1980).

\bibitem{ZelFro} A. I. Zel'nikov and V. P. Frolov, Sov. Phys. -JETP
  {\bf 55}, 191 (1982).

\bibitem{Lohi} D. Lohiya, J. Phys. A: Math. Gen. {\bf 15}, 1815 (1982).

\bibitem{LeLi1} B. L\'eaut\'e and B. Linet, J. Phys. A: Math. Gen. {\bf 15}, 1821 (1982).

\bibitem{LeLi2} B. L\'eaut\'e and B. Linet, Int. J. Theor. Phys. {\bf 22}, 67 (1983).

\bibitem{Beken2} J. D. Bekenstein, Phys. Rev. D {\bf 8}, 2333 (1973).

\bibitem{Wang} B. Wang, R. K. Su, P. K. N. Yu and E. C. M. Young, 
Phys. Rev. D {\bf 57}, 5284 (1998).

\end{thebibliography}
\end{document}